\documentclass[a4paper, 10pt, twocolumn]{article}
\usepackage{graphicx}
\usepackage{amssymb}
\usepackage{amsmath}
\usepackage{float}
\usepackage{url}
\usepackage{arydshln}
\usepackage{cite}
\usepackage{bm}
\begin{document}
\twocolumn
[
\begin{center}
{\huge Theoretical Analysis of SIRVVD Model to Provide Insight on the Target Rate of COVID-19/SARS-CoV-2 Vaccination in Japan}
\end{center}

\begin{center}
\begin{tabular}{cc}
{\Large \ } &\ \\
Yuto Omae$^1$, Makoto Sasaki$^1$, Jun Toyotani$^1$, Kazuyuki Hara$^1$, Hirotaka Takahashi$^2$
\\ \\
$^1$College of Industrial Technology, Nihon University
\\
$^2$Research Center for Space Science, Advanced Research Laboratories, Tokyo City University
{\Large \ } &\ \\
\end{tabular}
\end{center}
\vspace{5truemm}
{\bf Abstract}
The effectiveness of the first and second dose vaccinations are different for COVID-19; therefore, a susceptible-infected-recovered-vaccination1-vaccination2-death (SIRVVD) model that can represent the states of the first and second vaccination doses has been proposed.
By the previous study, we can carry out simulating the spread of infectious disease considering the effects of the first and second doses of the vaccination based on the SIRVVD model.
However, theoretical analysis of the SIRVVD Model is insufficient.
Therefore, we obtained an analytical expression of the infectious number, by treating the numbers of susceptible persons and vaccinated persons as parameters.
We used the solution to determine the target rate of the vaccination for decreasing the infection numbers of the COVID-19 Delta variant (B.1.617) in Japan. 
Further, we investigated the target vaccination rates for cases with strong or weak variants by comparison with the COVID-19 Delta variant (B.1.617). 
This study contributes to the mathematical development of the SIRVVD model and provides insight into the target rate of the vaccination to decrease the number of infections.
\vspace{15truemm}
]

\section{Introduction}
As of December 2021, the COVID-19 pandemic continues to be a global concern. Telework \cite{ref_tele}, lockdown \cite{ref_lockdown}, airport quarantine \cite{ref_air}, and tracing apps \cite{ref_app} are used as countermeasures for overcoming
COVID-19 spread.
The most important countermeasure is the COVID-19 vaccine (e.g., BNT162b (Pfizer) \cite{ref_pha}, mRNA-1273 (Moderna) \cite{ref_mode}, and ChAdOx1 (AstraZeneca) \cite{ref_ast}, and others). 
Mathematical simulation is a desirable approach to gain insight into the effect of vaccination on the infectious spread.
COVID-19 country-based simulations (Malaysia \cite{ref_vac_mal}, Saudi Arabia \cite{ref_vac_saudi}, Spain \cite{ref_vac_spa}, United States \cite{ref_vac_us}, and others) have already been reported.
A fast simulation method of differential equations is a susceptible-infected-recovered-vaccination (SIRV) model \cite{ref_SIRV1, ref_SIRV2, ref_SIRV3, ref_SIRV4, ref_SIRV5}.
However, these models can represent only the first dose of the vaccination.
For COVID-19, the effects of the first and second doses on infectious prevention are different; for the COVID-19 Delta variant (B.1.617), the effects of the first and second doses on infectious prevention are 35.6\% and 88.0\%, respectively \cite{pha}.
In this context, the SIRVVD model that represents the first and second doses of vaccination was proposed \cite{ref_omae_aims}.
However, theoretical analysis of the SIRVVD Model is insufficient.
Therefore, we find an analytical expression of the infectious number based on the SIRVVD model, by treating the numbers of susceptible persons and vaccinated persons as parameters.

\begin{figure}[bt]
 \centering
 \includegraphics[scale=0.18]{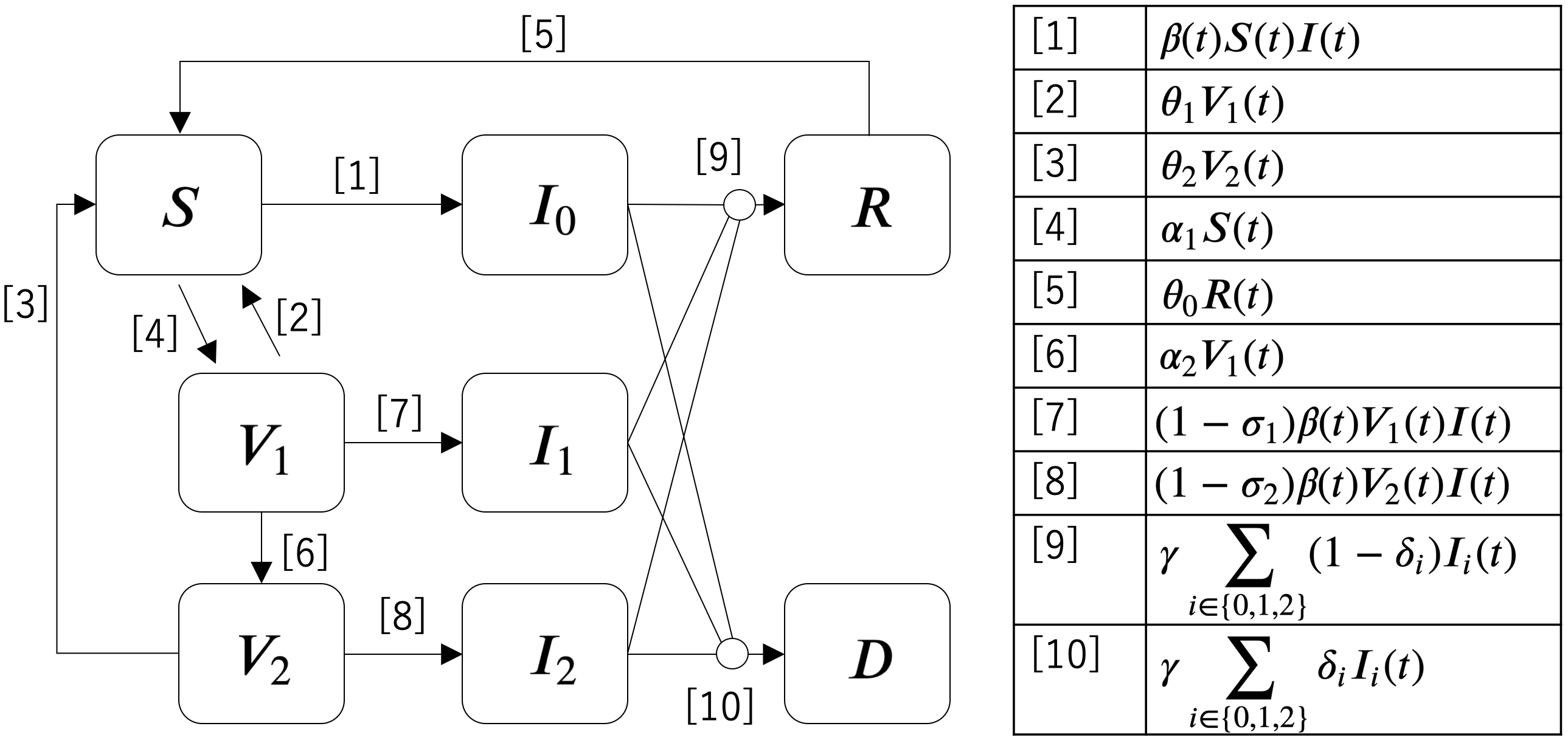}
 \caption{SIRVVD model \cite{ref_omae_aims}}
 \label{f0}
\end{figure}

In order to stop the spread of infection, it is important to achieve herd immunity through vaccination.
Some previous studies have reported various target values for achieving herd immunity through vaccination \cite{vobj_1, vobj_2, vobj_3, vobj_4, vobj_5}.
Because the target vaccination rate to achieve herd immunity depends on the country's lifestyle (wearing masks and social distancing) \cite{v_mask}, it is important to obtain target values of vaccination for each country.
Therefore, we calculated the target rate of the vaccinated persons in the scenario of spreading the COVID-19 Delta variant (B.1.617) in Japan based on the solution of infectious numbers of the SIRVVD model.
In addition, we investigated target rates in the case of becoming strong or weak variants compared with the COVID-19 Delta variant (B.1.617).

\section{SIRVVD model}
\subsection{Differential equations}
The SIRVVD model \cite{ref_omae_aims} for representing the first and second doses of vaccination is defined as
\begin{align}
\dot{S}(t) &= - \beta(t) S(t) I(t) + \theta_1 V_1(t) + \theta_2 V_2(t) \nonumber \\
               & ~~~~~ - \alpha_1 S(t) + \theta_0 R(t), \label{eq:s} \\
\dot{V_1}(t) &=  \alpha_1S(t) - \theta_1 V_1(t) - \alpha_2 V_1(t) \nonumber \\
               & ~~~~~ - (1-\sigma_1) \beta(t) V_1(t) I(t), \label{eq:v1} \\
\dot{V_2}(t) &=  \alpha_2 V_1(t) - (1-\sigma_2) \beta(t) V_2(t) I(t) \nonumber \\
               & ~~~~~ - \theta_2 V_2(t), \label{eq:v2}\\
\dot{I_0}(t) &=  \beta(t) S(t) I(t) - \gamma I_0 (t), \label{eq:i0}\\
\dot{I_1}(t) &=  (1-\sigma_1) \beta(t) V_1(t) I(t) - \gamma I_1 (t), \label{eq:i1} \\
\dot{I_2}(t) &=  (1-\sigma_2) \beta(t) V_2(t) I(t) - \gamma I_2 (t), \label{eq:i2}\\
\dot{R}(t) &=  \gamma \sum_{i \in\{0, 1, 2\}} (1-\delta_i) I_i(t) - \theta_0 R(t), \label{eq:r}\\
\dot{D}(t) &=  \gamma \sum_{i \in\{0, 1, 2\}} \delta_i  I_i(t), \label{eq:d}
\end{align}
where the dot notation represents the time derivative, i.e., $\dot{f}(t) := \mathrm{d}f(t)/\mathrm{d}t$.
Further, $S(t)$ represents the number of susceptible persons;
$I_{j \in \{0, 1, 2\}}(t)$ represents the number of infected persons of $j$ times the dose of the vaccination ($j = 0$ means no vaccination).
$V_{j \in \{1, 2\}}(t)$ represents the number of the vaccinated persons of $j$ times the dose of the vaccination.
$R(t)$ represents the number of recovered persons, and $D(t)$ represents the number of dead persons.
Moreover, the total infection number $I(t)$ is defined as
\begin{align}
I(t) = I_0 (t) + I_1(t) + I_2(t). \nonumber
\end{align}

Then, we describe the coefficient parameters as follows:
$\beta(t)$ represents infectivity, 
$\theta_{j \in \{0, 1, 2\}}^{-1}$ represents the average antibody period by the $j$ times the dose of vaccination ($j=0$ implies the natural infection).
$\alpha_{j \in \{1, 2\}}$ represents the transition rate to $V_j(t)$;
$\sigma_{j \in \{1, 2\}}$ represents the effectiveness of $j$ times the dose of vaccination (decreasing the value of infectious probability).
$\gamma^{-1}$ denotes the average infection period, and 
$\delta_{j \in \{0, 1, 2\}}$ denotes the fatality rate of the persons of $j$ times the dose of vaccination.
The states transitions are presented in Figure \ref{f0}.
The infection probability and fatality rate of the vaccinated persons $V_{j \in \{1, 2\}}$ decrease because of the vaccination effects.
Then, the total population $N$ is
\begin{align}
N &= S(t) + I(t) + V_1(t) + V_2(t) + R(t) + D(t), \nonumber \\
   \forall t &\in \bm{R}_0^{+}, \nonumber
\end{align}
where $\bm{R}_0^{+}$ represents the set of plus real numbers that include zero.

\subsection{Analytical expression of the infection number}
It is desirable that the function of the infectious number $I_{j \in \{0, 1, 2\}}(t)$ is clear to investigate the infectious spread.
However, it is difficult to obtain an analytical solution in the absence of any assumptions because $\dot{I}_j(t)$ is complex (including various functions). Therefore, we assume that $S(t), V_1(t), V_2(t)$, and $\beta(t)$ are not time-dependent functions but the constants of $S, V_1, V_2$, and $\beta$, respectively.
Then, the total number of persons that have an infectious possibility $N'$ is given by
\begin{align}
N' = S + V_1 + V_2. \label{s_del}
\end{align}
If the infectious number is sufficiently small, $N \simeq N'$.

$\dot{I}_j(t)$ defined by Equations (\ref{eq:i0})-(\ref{eq:i2}) can be simplified as
\begin{equation}
\begin{cases}
\dot{I}_0(t) = (\beta_0 - \gamma) I_0(t) + \beta_0 I_1(t) + \beta_0 I_2(t) \\
\dot{I}_1(t) = \beta_1 I_0(t) + (\beta_1 - \gamma) I_1(t) + \beta_1 I_2(t) \\
\dot{I}_2(t) = \beta_2 I_0(t) + \beta_2 I_1(t) + (\beta_2 - \gamma) I_2(t)
\end{cases}, \label{i_all}
\end{equation}
where 
\begin{align}
\beta_0 = \beta S, ~~ \beta_1 = \beta (1-\sigma_1) V_1, ~~ \beta_2 = \beta (1-\sigma_2) V_2. \nonumber
\end{align}
Herein, if we define
\begin{align}
\bm{I}(t) &= (I_0(t) ~~ I_1(t) ~~ I_2(t))^T, \nonumber
\\
\bm{A} &=
\begin{pmatrix}
\beta_0 - \gamma & \beta_0 & \beta_0 \\
\beta_1 & \beta_1-\gamma & \beta_1 \\
\beta_2 & \beta_2 & \beta_2-\gamma \\
\end{pmatrix}, \nonumber
\end{align}
then, Equation (\ref{i_all}) can be simplified into
\begin{equation}
\dot{\bm{I}}(t) = \bm{A} \bm{I}(t) \nonumber
\end{equation}
as the first-order separable simultaneous differential equations.
Here, we transform it into
\begin{equation}
\int \frac{1}{\bm{I}(t)} \mathrm{d}\bm{I}(t) = \bm{A}\int \mathrm{d}t, \nonumber
\end{equation}
and then, we can represent 
\begin{align}
\bm{I}(t) &= e^{\bm{A}t} \bm{I}(0) \nonumber \\
&= e^{\bm{P}\bm{U}\bm{P^{-1}}t} \bm{I}(0) \nonumber \\
&= \bm{P}e^{\bm{U}t}\bm{P^{-1}} \bm{I}(0) \nonumber \\
&= \bm{P} 
\begin{pmatrix}
e^{\lambda_0 t} & 0 & 0 \\
0 & e^{\lambda_1 t} & 0 \\
0 & 0 & e^{\lambda_2 t}
\end{pmatrix}
\bm{P^{-1}} \bm{I}(0), \label{i_stop}
\end{align}
where $\bm{U}$ represents the diagonal matrix of $\bm{A}$, and these relationships can be represented by
\begin{equation}
\bm{P^{-1}}\bm{A}\bm{P} = \bm{U} = \begin{pmatrix}
\lambda_0 & 0 & 0 \\
0 & \lambda_1 & 0 \\
0 & 0 & \lambda_2
\end{pmatrix}, \nonumber
\end{equation}
where $\bm{P}$ comprises the eigenvectors of the coefficient matrix $\bm{A}$, and $\lambda_{j \in \{0, 1, 2\}}$ represents the eigenvalues.
In other words, the solution $\bm{I}(t)$ is clear by calculating the eigenvectors and eigenvalues of the coefficient matrix $\bm{A}$.

To be clear, when $\lambda_{j \in \{0, 1, 2\}}$, let a $ 3\times3 $ unit matrix be denoted by $\bm{E}$; we calculate
\begin{align}
\mathrm{det} (\bm{A}-\lambda \bm{E}) &= (\beta_0 - \gamma - \lambda)(\beta_1 - \gamma - \lambda)(\beta_2 - \gamma - \lambda) \nonumber \\
& + 2 \beta_0 \beta_1 \beta_2 \nonumber \\
& - \beta_0 \beta_2 (\beta_1-\gamma - \lambda) \nonumber \\
& - \beta_1 \beta_2 (\beta_0-\gamma - \lambda) \nonumber \\
& - \beta_0 \beta_1 (\beta_2-\gamma - \lambda) \nonumber \\
& = (\gamma + \lambda)^2 (\beta_0 + \beta_1 + \beta_2 - \gamma - \lambda). \nonumber
\end{align}
In addition, the eigenvalues $\lambda$ are obtained by solving the eigenequation $\det(\bm{A} - \lambda \bm{E}) = 0$ as
\begin{align}
&\lambda_{j \in \{0, 1\}} = - \gamma \label{l1}, \\
&\lambda_2 = \beta_0 + \beta_1 + \beta_2 - \gamma \label{l2}.
\end{align}
From the eigenvalues, $\bm{P}$ comprises eigenvectors that can be represented as
\begin{align}
&\bm{P} =
\begin{pmatrix}
-1 & -1 & \frac{\beta_0}{\beta_2} \\
0 & 1 & \frac{\beta_1}{\beta_2} \\
1 & 0 & 1
\end{pmatrix}, \nonumber \\
&\bm{P}^{-1} = \frac{1}{B}
\begin{pmatrix}
-\beta_2 &  -\beta_2 & \beta_0+\beta_1 \\
-\beta_1 &  \beta_0 + \beta_2 & -\beta_1 \\
 \beta_2 &  \beta_2 & \beta_2 
\end{pmatrix}, \label{eqp}
\end{align}
where
\begin{align}
B = \beta_0 + \beta_1 + \beta_2, ~~ B \neq 0. \nonumber
\end{align}

The solution $\bm{I}(t)$ of the differential equation $\dot{\bm{I}}(t)$ can be represented by substituting Equations (\ref{l1})-(\ref{eqp}) into Equation (\ref{i_stop}).
Then, solutions $I_{j \in \{0, 1, 2\}}$ are represented as 
\begin{equation}
I_j(t) = X_j e^{\bigl\{ \beta \bigl( S+\sum_{i \in \{1, 2\}}(1-\sigma_i)V_i\bigr) -\gamma \bigr\}t} + Y_j e^{-\gamma t} \label{i0_sol}.
\end{equation}
The coefficients are given by
\begin{align}
X_0 &= \frac{S I(0)}{S+\sum_{i \in \{1, 2\}}(1-\sigma_i)V_i}, \nonumber \\
X_l &= \frac{(1-\sigma_l)V_l I(0)}{S+\sum_{i \in \{1, 2\}}(1-\sigma_l)V_l}, \nonumber \\
Y_0 &=  \frac{\sum_{i \in \{1, 2\}}\{ (1-\sigma_i)V_i I_0(0) - SI_i(0) \}}{S+\sum_{i \in \{1, 2\}}(1-\sigma_i)V_i}, \nonumber \\
Y_l &=  \frac{\bigl( S + (1-\sigma_{k})V_{k} \bigr) I_l(0) - (1-\sigma_l)V_l \bigl(I_0(0) + I_{k}(0)\bigr) }{S+\sum_{i \in \{1, 2\}}(1-\sigma_i)V_i}, \nonumber
\end{align}
where, 
\begin{align}
& l \in L = \{1, 2\}, \nonumber \\
&k \in L \setminus \{l\}, \nonumber \\
&S+\sum_{i \in \{1, 2\}}(1-\sigma_i)V_i \neq 0. \nonumber
 \end{align}
These equations require constant values of $S(t), V_1(t)$, and $V_2(t)$;
thus, these equations are not suitable for long-term simulations.

\subsection{Target rate of the vaccination}
$\gamma$ represents the inverse value of the average infection period, and therefore, $\gamma > 0$.
Thus, 
\begin{equation}
\lim_{t \to \infty} e^{-\gamma t} = 0. \nonumber
\end{equation}
Further, because the coefficients of Equation (\ref{i0_sol}) are constants, each solution $I_{j \in \{0, 1, 2\}}(t)$ can be approximated by
\begin{align}
I_{j \in \{0, 1, 2\}} &\sim e^{\bigl\{ \beta \bigl( S+\sum_{i \in \{1, 2\}}(1-\sigma_i)V_i\bigr) -\gamma \bigr\}t} \nonumber \\
&= e^{\bigl\{ \beta \bigl( N' - \sum_{i \in \{1, 2\}} \sigma_i V_i \bigr) -\gamma \bigr\}t}. \nonumber
\end{align}
$S$ was deleted by Equation (\ref{s_del}). Therefore, we get the following theorems:
\begin{align}
&\beta(N' - \sum_{i \in \{1, 2\}} \sigma_i V_i) - \gamma > 0 \Rightarrow \mbox{$I_{j}(t)$ increases,} \label{cond0}\\
&\beta(N' - \sum_{i \in \{1, 2\}} \sigma_i V_i) - \gamma = 0 \Rightarrow \mbox{$I_{j}(t)$ does not change,}  \label{cond1}\\
&\beta(N' - \sum_{i \in \{1, 2\}} \sigma_i V_i) - \gamma < 0 \Rightarrow \mbox{$I_{j}(t)$ decreases,}\label{cond2}
\end{align}
where, $j \in \{0, 1, 2\}$.
The infectious spread stops by creating a scenario as indicated in Equation (\ref{cond1}). 
We collected terms on virus parameters to the right side and vaccination parameters on the left side to clarify this scenario.
As a result, Equation (\ref{cond1}) can be simplified into
\begin{equation}
N' - \sigma_1 V_1 - \sigma_2 V_2 = \gamma \beta^{-1}. \nonumber
\end{equation}
Here, $N'$ denotes the total number of persons with a probability of infection;
$\sigma_{j \in \{1, 2\}}$ represents the effectivity of the vaccination;
and $V_{j \in \{1, 2\}}$ represens the number of vaccinated persons.
Thus, the left side represents the total number of people who have the probability of infection considering the vaccination (the left side is small when using highly effective vaccination).
The longer the infection period and the higher the infectivity, the smaller is the right term because $\gamma^{-1}$ represents the average infection period and $\beta$ represents the infectivity.
In addition, in the case a virus has a long infection period and high infectivity, a highly effective vaccination is required to stop the spread of infection.

Here, let $V_2^\mathrm{obj.}$ denote the target number of the second dose of vaccination required to stop the spread of infection. 
We can obtain $V_2^\mathrm{obj.}$ by solving Equation (\ref{cond1}) for $V_2$ as
\begin{equation}
V_2^\mathrm{obj.} = \frac{N' - \sigma_1 V_1 - \gamma \beta^{-1}}{\sigma_2}.  \label{obj}
\end{equation}
Further, the target rate of the second dose of vaccination required to stop the infection spread $P^\mathrm{obj.}_2$ is 
\begin{equation}
P^\mathrm{obj.}_2 = V_2^\mathrm{obj.}/N'. \nonumber
\end{equation}

\subsection{Limitation of the vaccination}
We consider the condition of not stopping the infection spread even if all persons receive the second dose of vaccination.
We obtain the following theorem because this scenario would mean substituting $V_2=N', V_1=0$ for Eq.(\ref{cond0}).
\begin{align}
&\sigma_2 < 1 - \frac{\gamma}{\beta N'} \Rightarrow \nonumber \\
&\mbox{$I_{j \in \{0, 1, 2\}}(t)$ increases when $V_1=0$ and $V_2=N'$}. \label{cond3}
\end{align}
Even if all persons receive the second dose of vaccination, the infection number continues to increase when the effectivity of the second dose of the vaccination $\sigma_2$ is insufficient.
In contrast, the condition of Equation (\ref{cond3}) is not satisfied when the vaccination that has perfect prevention (i.e., $\sigma_2=1$) because $N'>0, \beta>0$, and $\gamma>0$.
In other words, the spread of infection stops when developing a vaccination that has a perfect prevention effect (i.e., $\sigma_2=1$) and getting a vaccination to all persons (i.e., $V_2=N'$).
However, it is important to know the condition of Equation (\ref{cond3}) because achieving this is difficult.

\subsection{Conditions under which the vaccine is not required}
We consider conditions under which the vaccine is not required; this implies substituting $V_2 = 0, V_1 = 0$ for Equation (\ref{cond2}).
We can obtain a simplified equation as
\begin{align}
&0 > 1-\frac{\gamma}{\beta N'} \Rightarrow \nonumber \\
&\mbox{when $V_{j\in\{1, 2\}}=0$, $I_{j \in \{0, 1, 2\}}$ decreases.} \label{cond4}
\end{align}

\begin{figure}[t]
 \centering
 \includegraphics[scale=0.52]{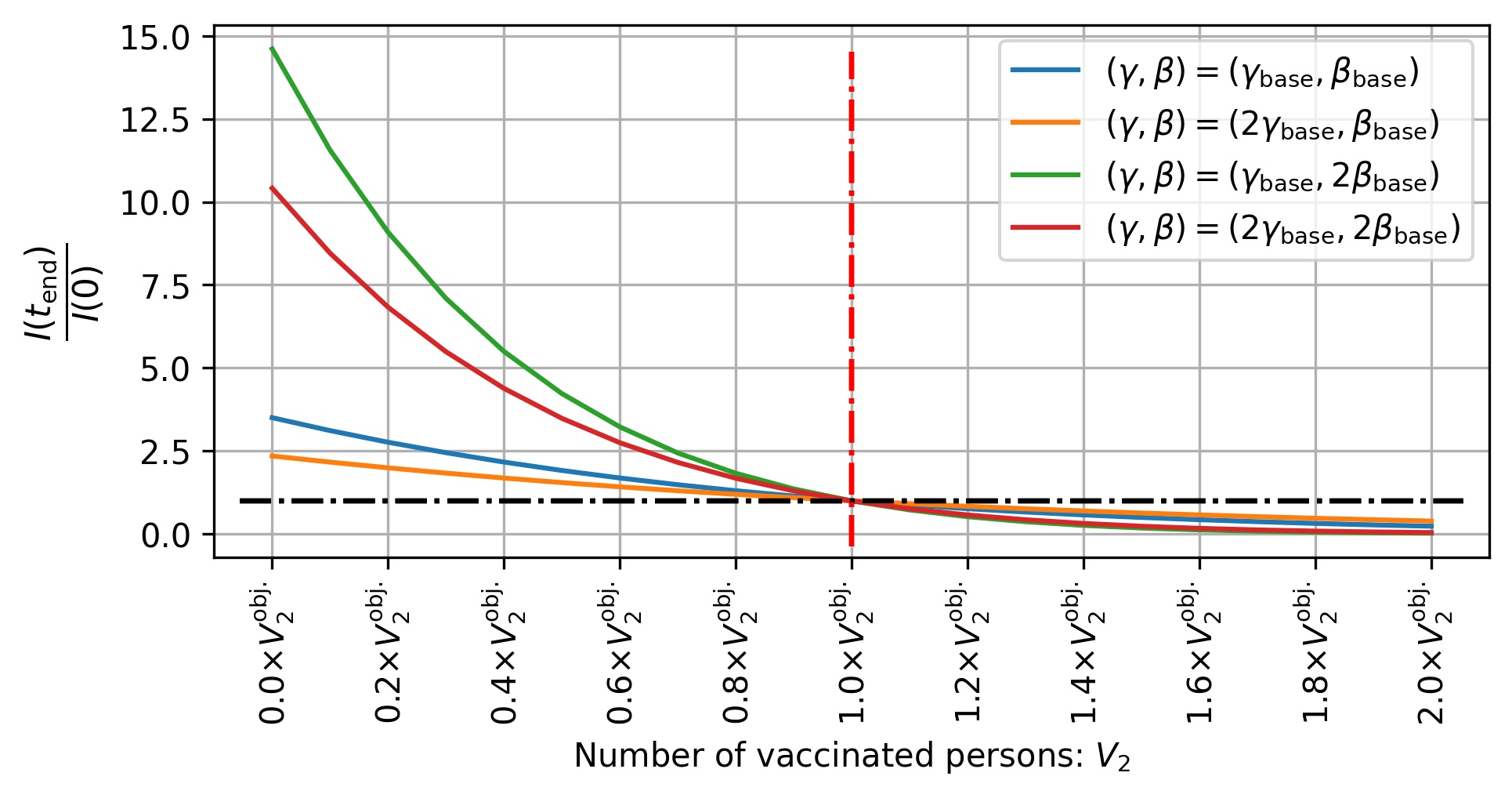}
 \caption{Relationship between the number of people who get the second dose of vaccination $V_2$ and the infection spread ($V_1 = 0$).}
 \label{f1}
\end{figure}

\subsection{Validity of the target number $V_2^\mathrm{obj.}$}
Here, we check whether the infection spread stops by satisfying $V_2^\mathrm{obj.}$.
Therefore, we conduct the SIRVVD model-based simulations (i.e., Equations (\ref{eq:s}) - (\ref{eq:d})) assuming a simple scenario.
The fixed parameters are 
\begin{itemize}
\setlength{\itemsep}{0pt}
\item Simulation period $t_\mathrm{end}=10$ [days]
\item Total population $N=2 \times 10^6$
\item Initial infection numbers of $j$ times vaccinated persons $I_{j \in \{0, 1, 2\}}(0)=10^3$
\item Initial dead persons $D(0)=0$
\item Initial recovered persons $R(0)=0$
\item Initial susceptible persons $S(0)=N-\sum_j I_j(0)-D(0)-R(0)-\sum_j V_j(0)$
\item Effectivity of the first dose of vaccination $\sigma_1=0.50$
\item Effectivity of the second dose of vaccination $\sigma_2=0.90$
\item Vaccination rate $\alpha_{j \in \{1, 2\}} = 0$
\item Average period of antibody $\theta_{j \in \{0, 1, 2\}}^{-1} \rightarrow \infty$ (i.e., $\theta_{j \in \{0, 1, 2\}} = 0$)
\item Fatality rate $\delta_{j \in \{0, 1, 2\}} = 0$
\end{itemize}
The parameters validated by multiple values are 
\begin{itemize}
\setlength{\itemsep}{0pt}
\item Removal rate $\gamma \in \{\gamma_\mathrm{base}, 2\times \gamma_\mathrm{base}\}$ ($\gamma_\mathrm{base}=0.05$)
\item Infectivity $\beta \in \{\beta_\mathrm{base}, 2\times \beta_\mathrm{base}\}$ ($\beta_\mathrm{base}=10^{-7}$)
\item Initial number of vaccinated persons (first dose) $V_1 \in \{0, N/2\}$
\end{itemize}
We adopted $V_2 \in \{0 \times V_2^\mathrm{obj.}, 0.2 \times V_2^\mathrm{obj.}, \cdots, 2.0 \times V_2^\mathrm{obj.}\}$ over the initial number of persons who get the second dose of vaccination $V_2$ to verify the cases that satisfy $V_2^\mathrm{obj.}$.

\begin{figure}[t]
 \centering
 \includegraphics[scale=0.52]{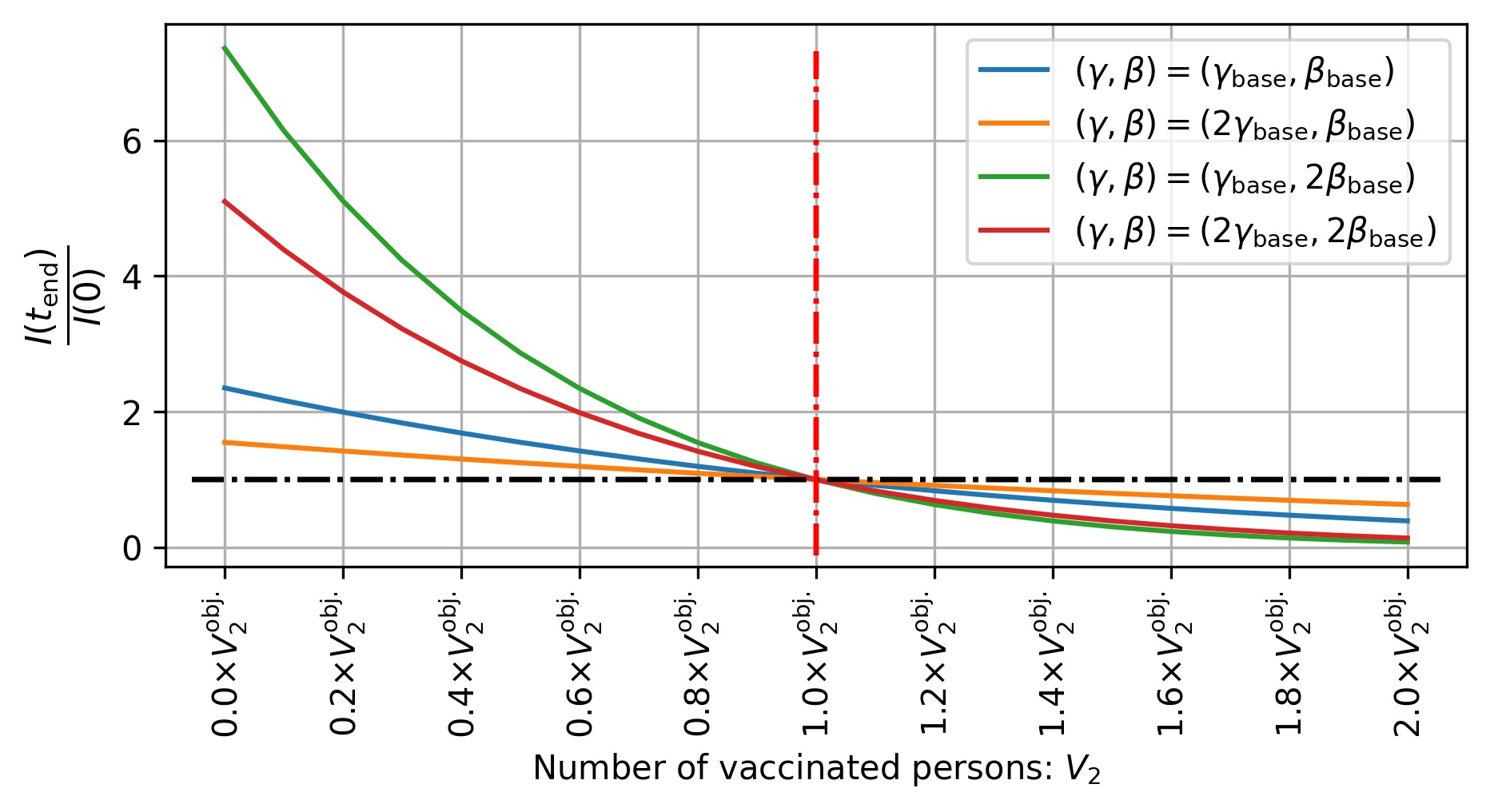}
 \caption{Relationship between the number of persons who get the second dose of vaccination $V_2$ and the infection spread ($V_1 = N/2$).}
 \label{f2}
\end{figure}

The simulation results for cases $V_1 = 0$ and $V_1 = N/2$ are shown in Figures \ref{f1} and \ref{f2}, respectively.
The horizontal axis represents $V_2$; the vertical axis, $I(t_\mathrm{end})/I(0)$.
In addition, the black dashed line represents $I(t_\mathrm{end})/I(0)=1$; the red dashed line, $V_2=V_2^\mathrm{obj.}$.
$I(t_\mathrm{end})/I(0)>1$ indicates that the infection number increases, and $I(t_\mathrm{end})/I(0)<1$ implies that it decreases.
In all cases, when $V_2 = V_2^\mathrm{obj.}$, $I(t_\mathrm{end})/I(0) = 1$.
Further, the infection number increases ($I(t_\mathrm{end})/I(0)>1$) when the number of vaccinated persons is low ($V_2 < V_2^\mathrm{obj.}$).
In contrast, the infection number decreases ($I(t_\mathrm{end})/I(0)<1$) when the number of vaccinated persons is sufficient ($V_2 > V_2^\mathrm{obj.}$).
Therefore, the target number of the second dose of vaccination, $V_2^\mathrm{obj.}$ based on Equation (\ref{obj}) can be considered reliable.

\section{Target rate $P^\mathrm{obj.}_2$ under the scenario of the COVID-19 Delta variant (B.1.617) in Japan}
We calculate the target rate $P^\mathrm{obj.}_2$ of the second dose of vaccination assuming the case of the COVID-19 Delta variant (B.1.617) in Japan.

\subsection{Infectivity of the COVID-19 Delta variant (B.1.617) in Japan}
Let $\beta_\mathrm{delta}$ denote the infectivity of the COVID-19 Delta variant (B.1.617) in Japan; we estimate this value.
According to previous research \cite{ref_omae_aims2}, infectivity $\beta$ can be represented as
\begin{equation}
\beta = r \gamma S^{-1}, \label{r2b}
\end{equation}
where $r$ denotes the effective reproduction number.
Further, $\beta$ calculated by Equation (\ref{r2b}) includes the effects of various countermeasures since the effective reproduction number includes these effects.
We can thus obtain the infectivity $\beta_\mathrm{delta}$ if we know the effective reproduction number of the delta variant (B.1.617) in Japan.
According to the COVID-19 advisory board for the Japanese government \cite{delta_rate}, after August 2021, all infected people had the delta variant (B.1.617).
Thus, the effective reproduction number during this period can be attributed to the COVID-19 Delta variant (B.1.617).
However, on August 31, 2021, the rates of people who received the first and second vaccination doses were 11\% and 46\%, respectively (total rate: 57\%) \cite{vrate}.
Therefore, we consider that the infectivity of the Delta variant (B.1.617) is underestimated owing the effect of the vaccination when considering the newer cases of infection during this period.

Thus, we estimated the infectivity of the Delta variant (B.1.617) in Japan using data from the period when the vaccination was not started (before February 2021).
According to the COVID-19 advisory board for the Japanese government \cite{delta}, the Delta variant (period: August 2021) had a 1.9 times higher infection rate compared to the that of the regular COVID-19 virus (period: December 2020) in Japan because no people were vaccinated in 
December 2020.
Thus, we consider the infectivity of the Delta variant (B.1.617) as the 1.9 times the infectivity of this period.
We assume 
\begin{equation}
\beta_\mathrm{delta} = 1.9 \beta_\mathrm{regular}, \label{reg2del}
\end{equation}
where $\beta_\mathrm{delta}$ and $\beta_\mathrm{regular}$ represent the infectivity of the Delta variant (B.1.617) and regular COVID-19 virus, respectively.

The maximum value of the effective reproduction number in December 2020 in Japan was $r = 1.18$ \cite{toyo}.
Kobayashi et al. estimated $0.17$ as the $\gamma$ parameter of COVID-19 in Japan from $0.13 $ to $ 0.17 $ \cite{kobayashi}.
We adopted $\gamma = 0.13$ because we consider optimism to be undesirable, and this leads to a long infection period.
The total population in Japan is approximately $S=1.2 \times 10^8$ persons.
We obtain $\beta_\mathrm{regular} \simeq 1.28 \times 10^{-9}$ as the infectivity of the COVID-19 regular virus by substituting these parameters in Equation (\ref{r2b}).
Moreover, the infectivity of the Delta variant (B.1.617) is $\beta_\mathrm{delta} \simeq 2.43 \times 10^{-9 }$ by substituting $\beta_\mathrm{regular}$ for Equation (\ref{reg2del}) and by substituting $\beta_\mathrm{regular }$.

\begin{table}[bt]
\caption{Target rates of vaccinated persons to stop infection spread reported by other researches.}
  \label{tab1}
  \centering
  \begin{tabular}{lccc}\hline
    \multicolumn{1}{c}{Paper} & Area & Effect* & Target rate  \\ \hline
    Agarwal \cite{vobj_1} & World & 82.5\% & 45 - 60 \% \\
    Gumel \cite{vobj_2} & US & 70\% & 80 \% \\
    Gumel \cite{vobj_2} & US & 95\% & 60 \% \\
    Kadkhoda \cite{vobj_3} & - & 100\% & 62 - 72 \% \\
    Kadkhoda \cite{vobj_3} & - & 95\% & 63 - 76 \% \\
    Fontanet \cite{vobj_4} & France & - & 67 \% \\
    Liu \cite{vobj_5} & China & 90\% &  83 - 92\% \\ \hdashline
    Ours & Japan** & 88\% & 63\% \\
    \hline 
    \multicolumn{4}{r}{*: Prevention effect of vaccine.} \\
    \multicolumn{4}{r}{**: The case of lifestyle on Dec. 2020 in Japan.}
  \end{tabular}
\end{table}

\subsection{Effectiveness of the BNT162b (Pfizer) vaccination }
In a previous study, Bernal et al. \cite{pha} reported the effect of the BNT162b (Pfizer) vaccine on reducing the probability of infection with the COVID-19 Delta variant (B.1.617).
The effect of the first dose was 35.6\% (22.7$\sim$46.4\%), and that of the second dose was 88.0\% (85.3$\sim$90.1\%).
Thus, we adopted $\sigma_1=0.356$ and $\sigma_2=0.880$.

\begin{figure*}[t]
 \centering
 \includegraphics[scale=0.7]{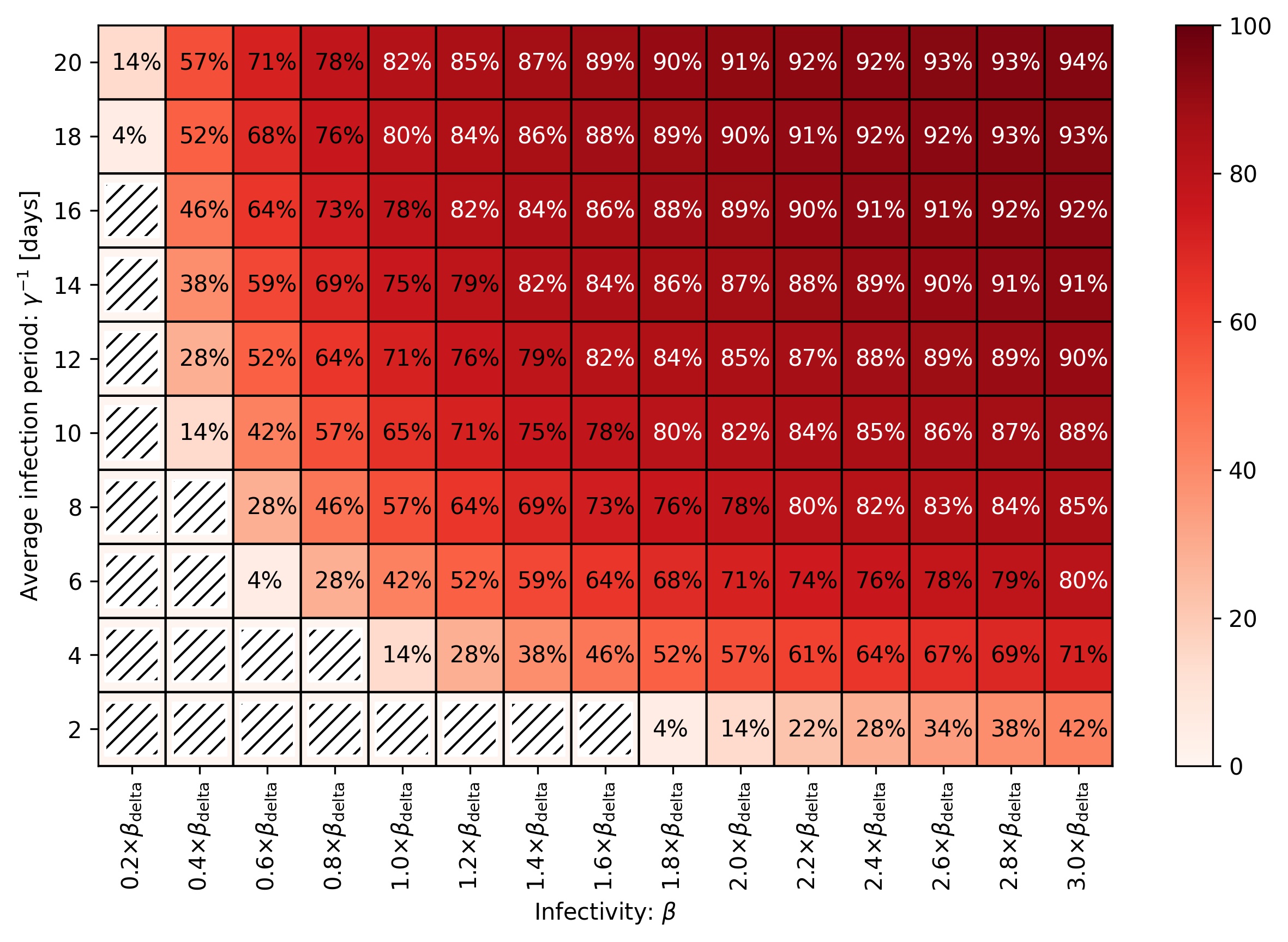}
 \caption{
 Target rates $P^\mathrm{obj.}_2$ ($\sigma_2=1$).
 White hatchings on the left lower side indicate the case of satisfying Equation (\ref{cond4}).
}
 \label{f3}
\end{figure*}

\begin{figure*}[t]
 \centering
 \includegraphics[scale=0.7]{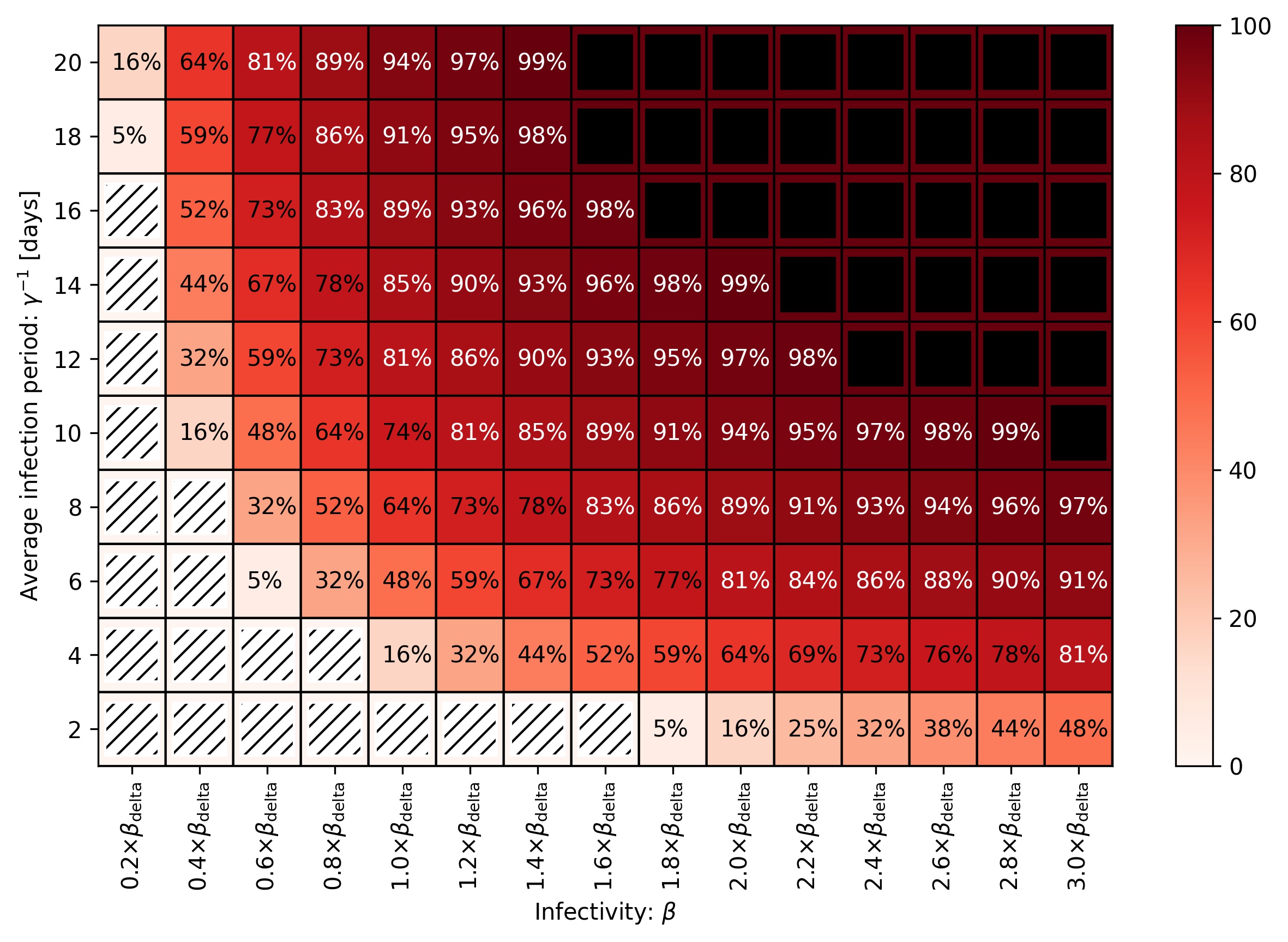}
 \caption{
  Target rates $P^\mathrm{obj.}_2$ ($\sigma_2=0.88$).
  White hatchings on the left lower side indicate the case of satisfying Equation (\ref{cond4}).
  Painted black markers represent the results satisfying Equation (\ref{cond3}).
}
 \label{f4}
\end{figure*}

\subsection{Target number $V^\mathrm{obj.}_2$ and target rate $P^\mathrm{obj.}_2$ in Japan}
We consider the target number $V^\mathrm{obj.}_2$ and target rate $P^\mathrm{obj.}_2$ in Japan for the infectious spread of the Delta variant (B.1.617).
We assume that all persons who received the first dose will receive the second dose, i.e., $V_1 = 0$.
Further, the total population in Japan is about $1.2 \times 10^8$.
$V^\mathrm{obj.}_2$ can be represented by substituting these values for Equation (\ref{obj}) as
\begin{align}
V_2^\mathrm{obj.} &= \frac{N' - \sigma_1 V_1 - \gamma \beta^{-1}}{\sigma_2} \nonumber \\
&= \frac{1.2 \times 10^8 - 0.356 \times 0 - 0.13 \times (2.43\times 10^{-9})^{-1}}{0.880} \nonumber \\
&\simeq 7554 \times 10^4. \nonumber
\end{align}
Thus, the spread of infection will stop when $7554 \times 10^4$ persons receive the second dose of the vaccination. 
The target rate is $P_2^\mathrm{obj.} =V _2^\mathrm{obj.}/N' = 0.6295$.
A vaccination rate of approximately 63\% is required for the total population in Japan to stop the spread of infection.
The target rate of 63\% is for the case of the lifestyle of December 2020 in Japan.
If we obtain the lifestyle values before the appearance of COVID-19 (before December 2019), a target rate of about 63\% will appear to be insufficient.

Snehota et al. \cite{v_accept} reported that about 72.5\% of the population in various countries have taken the COVID-19 vaccine.
We believe that the target vaccination rate can be achieved in Japan because the target rate $P^\mathrm{obj.}_2 = 0.63$ is lower than the reported global value of 72.5\%.
In Japan, on December 27, 2021, 78\% of the total population received the second dose of vaccination \cite{vrate}, and the infection number decreased (e.g., the average number of new cases of infection per day from November 1, 2021, to December 31, 2021 was lower than 3 persons per million persons \cite{vcases}.).

\subsection{Comparison with other studies}
The target rates of the vaccinated persons reported by other researchers are listed in Table \ref{tab1}.
These are target rates for achieving herd immunity to stop infection spread. We check the various values in this study.

One reason for the different values is that the target rate is dependent on the lifestyle of each country. For example, Shen et al. \cite{v_mask} reported that the target rate of vaccination is dependent on wearing masks and social distancing.

The target rate of vaccinated persons to stop the infection spread is not the same because the lifestyles of each country are not the same.
Thus, it is necessary to calculate the target rate of each country.
$P^\mathrm{obj.}_2 = 0.63$ represents the target rate for the lifestyle in December 2020 in Japan.

\begin{figure*}[t]
 \centering
 \includegraphics[scale=0.7]{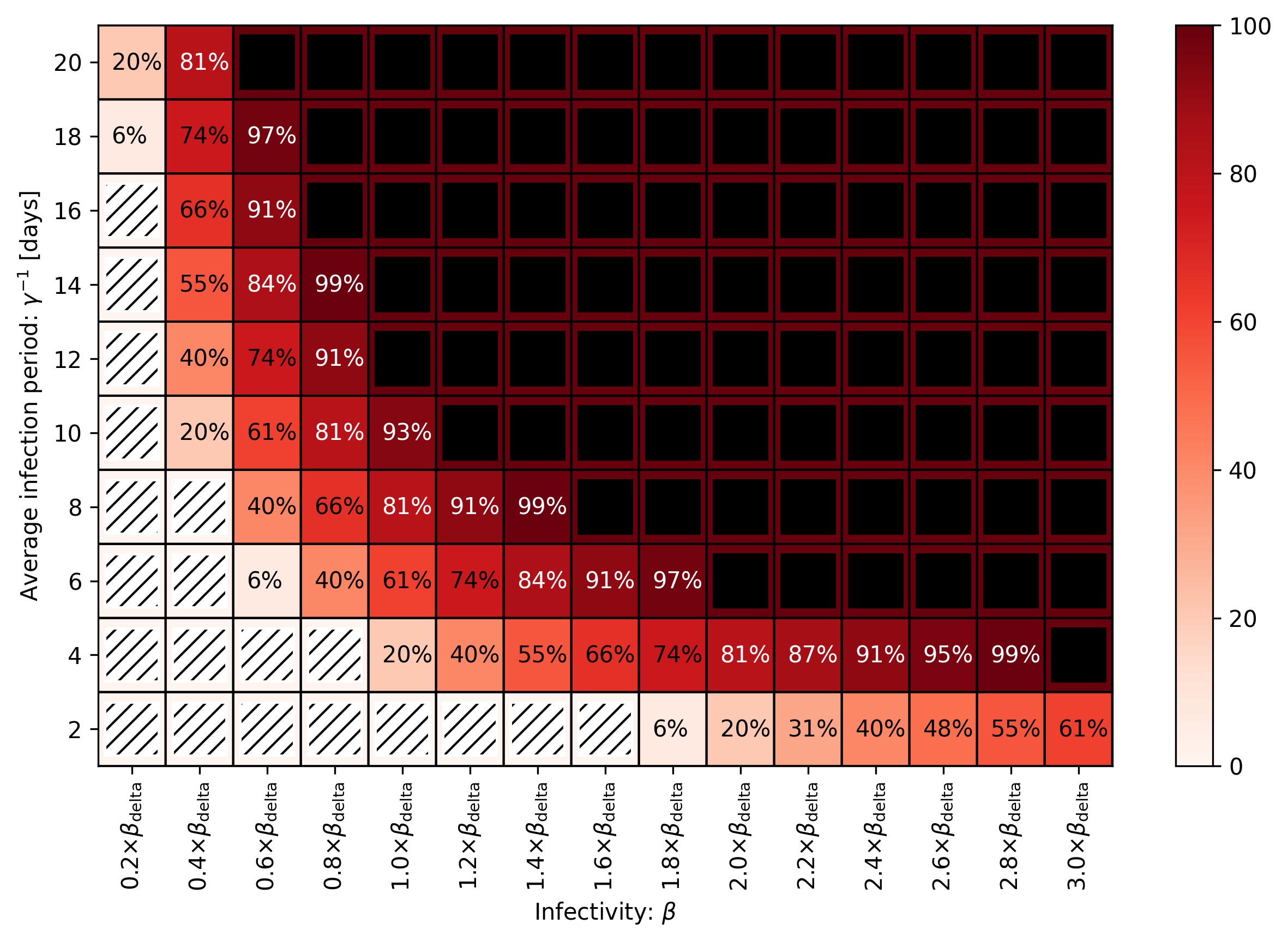}
 \caption{
  Target rates $P^\mathrm{obj.}_2$ ($\sigma_2=0.70$).
  White hatchings on the left lower side indicate the case of satisfying Equation (\ref{cond4}).
  Painted black markers represent the results satisfying Equation (\ref{cond3}).
}
 \label{f5}
\end{figure*}

\section{Cases of strong or weak new variants}
In the near future, there is a possibility of new variants occurring; for example, the COVID-19 Omicron variant (B.1.1.529) \cite{omic}.
Thus, we consider the cases of strong or weak new variant comparisons with the Delta variant (B.1.617).
We consider the following conditions: 
\begin{itemize}
\setlength{\itemsep}{0pt}
\item Changing the average infection period $\gamma^{-1}$
\item Changing infectivity $\beta$
\item Changing vaccination effectivity $\sigma_2$
\end{itemize}

For the simulation parameters, we considered the average infection periods ranging from 2 - 20 days, and the infectivity ranging from 0.2 - 3 times.
In addition, we consider three patterns as vaccination effectiveness: the case of perfect prevention ($\sigma_2 = 1$); the case of BNT162b (Pfizer) effectivity ($\sigma_2=0.88$ \cite{pha}); and the case of a lower effectivity comparison with BNT162b (Pfizer) ($\sigma_2=0.70$). 
Thus, we adopted the combinations of the following parameters
\begin{align}
&\gamma^{-1} \in \{2, 4, \cdots, 20\},\nonumber \\
&\beta \in \{0.2 \times \beta_\mathrm{delta}, ~ 0.4 \times \beta_\mathrm{delta}, ~ \cdots, ~ 3.0 \times \beta_\mathrm{delta}\},\nonumber \\
&\sigma_2 \in \{1, ~ 0.88, ~ 0.70\}, \nonumber
\end{align}
and calculated the target rate $P^\mathrm{obj.}_2$ to stop the infection spread.

The target rates $P^\mathrm{obj.}_2$ in the case of perfect prevention ($\sigma_2=1$) are shown in Figure \ref{f3}.
The vertical axis represents the average infection period $\gamma^{-1}$ [day], and the horizontal axis represents the infectivity $\beta$ ($1.0 \times \beta_\mathrm{delta}$ represents the Delta variant infectivity).
The percentage values are $P^\mathrm{obj.}_2$ and the white hatchings on the left lower side indicate the case of satisfying Equation (\ref{cond4}), i.e., the vaccination is not required to stop the spread of infection.
This result suggests that the spread of infection will stop if almost all persons receive the second dose of vaccination even if the new variant has a higher infectivity than the Delta variant (B.1.617),.

However, the perfect prevention vaccine is not realistic.
Therefore, we calculate the case of $\sigma_2=0.88$, as shown in Figure \ref{f4}.
Painted black markers represent the results satisfying Equation (\ref{cond3}), i.e., even if all persons receive the second dose of vaccination, the infection spread does not stop.
This scenario occurs in the cases of long infection periods and high infectivity. For such a scenario, we consider not only vaccinations but also lockdowns, such as stay-at-home orders.

Finally, the case of a new variant that leads to a decrease in the vaccination effect ($\sigma_2=0.70$) is shown in Figure \ref{f5}.
There are many cases wherein only the vaccination cannot stop the spread of the infection caused by a new strong variant. 
The results indicate that it is important to develop more high-effect vaccines to overcome the new strong variant.

\section{Conclusion}
We found the analytical expressions of $I_{j\in\{0, 1, 2\}}(t)$ of the differential equations $\dot{I}_{j\in\{0, 1, 2\}}(t)$ of the SIRVVD model.
Furthermore, we proposed a method to determine $P^\mathrm{obj.}_2$, which is the target rate of the vaccination required to stop the spread of infection.
Assuming the COVID-19 Delta variant (B.1.617) in Japan, we estimated $P^\mathrm{obj.}_2$ to be approximately 63\%.
Further, we calculated the target rate of the vaccination by assuming a new strong or weak variant (Figures \ref{f3}, \ref{f4}, and \ref{f5}).
We consider that these values are important to control the infection spread.

We did not consider the validity term of the vaccination.
The vaccination effect decreases based on elapsed time \cite{omic}.
In future studies, we plan to investigate vaccination strategies that consider the validity of the vaccination.


\begin{thebibliography}{99}
\bibitem{ref_tele}
G. Buomprisco, S. Ricci, R. Perri, S. De Sio, Health and telework: New challenges after the COVID-19 pandemic, Eur. J. Environ. Public Health, 5 (2021), em0073. doi: 10.21601/ejeph/9705.

\bibitem{ref_lockdown}
H. Lau, V. Khosrawipour, P. Kocbach, A. Mikolajczyk, J. Schubert, J. Bania, et al.,
The positive impact of lockdown in Wuhan on containing the COVID-19 outbreak in China,
J. Travel Med., (2020). doi: 10.1093/jtm/taaa037.

\bibitem{ref_air}
T. Sekizuka, K. Itokawa, K. Yatsu, R. Tanaka, M. Hashino, T. Kawano-Sugaya, et al., COVID-19 genome surveillance at international airport quarantine stations in Japan, J. Travel Med., 28 (2021), taaa217. doi: 10.1093/jtm/taaa217.

\bibitem{ref_app}
N. Ahmed, R. A. Michelin, W. Xue, S. Ruj, R. Malaney, S. S. Kanhere, et al., A survey of COVID-19 contact tracing apps, IEEE Access, 8 (2020), 134577--134601. doi: 10.1109/ACCESS.2020.3010226.

\bibitem{ref_pha}
A. B. Vogel, I. Kanevsky, Y. Che, K. A. Swanson, A. Muik, M. Vormehr, et al., BNT162b vaccines protect rhesus macaques from SARS-CoV-2, 
Nature, 592 (2021), 283--289. doi: 10.1038/s41586-021-03275-y.

\bibitem{ref_mode}
N. Doria-Rose, M. S. Suthar, 
M. Makowski, S. O'Connell, 
A. B. McDermott, 
B. Flach, et. al., Antibody persistence 6 months after the second dose of mRNA-1273 vaccine for Covid-19. New Eng. J. Med., 384 (23) (2021), 2259--2261. doi: 10.1056/NEJMc2103916.

\bibitem{ref_ast}
M. Scully, D. Singh, R. Lown, A. Poles, T. Solomon, 
M. Levi, et. al., Pathologic antibodies to platelet factor 4 after vaccination with ChAdOx1 nCoV-19, New Eng. J. Med., 384 (23) (2021), 2202--2211. doi: 10.1056/NEJMoa2105385.

\bibitem{ref_vac_mal} 
W. K. Wong, F. H. Juwono, T. H. Chua, SIR simulation of covid-19 pandemic in Malaysia: Will the vaccination program be effective?  arXiv preprint, arXiv: 2101.07494 (2021). \url{https://arxiv.org/abs/2101.07494}

\bibitem{ref_vac_saudi} 
R. Ghostine M. Gharamti, S. Hassrouny, I. Hoteit,
An extended SEIR model with vaccination for forecasting the COVID-19 pandemic in Saudi Arabia using an ensemble Kalman filter, Mathematics, 9 (6) (2021), 636. doi: 10.3390/math9060636.

\bibitem{ref_vac_spa} 
P. Kumar, V. S. Erturk, M. Murillo-Arcila, 
A new fractional mathematical model of COVID-19 with the availability of vaccines. Results Phys., 24 (2021) 104213. doi: 10.1016/j.rinp.2021.104213.

\bibitem{ref_vac_us} 
J. Li, P. Giabbanelli, Returning to a normal life via COVID-19 vaccines in the United States: A large-scale agent-based simulation study, JMIR Med. Inform., 9(4) (2021), e27419. doi: 10.2196/27419.

\bibitem{ref_SIRV1}
X. Meng, Z. Cai, H. Dui, H. Cao, Vaccination strategy analysis with SIRV epidemic model based on scale-free networks with tunable clustering, IOP Conf. Ser.: Mater. Sci. Eng., 1043 (2021), 032012. doi: 10.1088/1757-899X/1043/3/032012.

\bibitem{ref_SIRV2}
R. Rifhat, Z. Teng, C. Wang, 
Extinction and persistence of a stochastic SIRV epidemic model with a nonlinear incidence rate. Adv. Differ. Equ., (2021) 1--21. doi: 10.1186/s13662-021-03347-3.

\bibitem{ref_SIRV3}
M. Ishikawa, Optimal strategies for vaccination using the stochastic SIRV model,
Trans. Inst. Syst., Control, Inf. Eng., 25 (2012), 343--348. doi: 10.5687/iscie.25.343.

\bibitem{ref_SIRV4}
J. Farooq, M. A. Bazaz, A novel adaptive deep learning model for Covid-19 with a focus on mortality reduction strategies. Chaos, Solitons, Fractals, 138  (2020), 110148. doi: 10.1016/j.chaos.2020.110148.

\bibitem{ref_SIRV5}
M. O. Oke, O. M. Ogunmiloro, C. T. Akinwumi, R. A. Raji, Mathematical modeling and stability analysis of a SIRV epidemic model with nonlinear force of infection and treatment. Commun. Math. Appl., 10 (4) (2019), 717--731.

\bibitem{pha}
J. Lopez Bernal, 
N. Andrews, 
C. Gower, 
E. Gallagher, 
R. Simmons, 
S. Thelwall,
et. al.,
Effectiveness of Covid-19 vaccines against the B.1.617.2 (delta) variant, N. Engl. J. Med., (2021), 585--594. doi: 10.1056/NEJMoa2108891.

\bibitem{ref_omae_aims}
Y. Omae, Y. Kakimoto, M. Sasaki, J. Toyotani, K. Hara, Y. Gon, et al., SIRVVD model-based verification of the effect of first and second doses of COVID-19/SARS-CoV-2 vaccination in Japan, Math. Biosci. Eng., 19 (1) (2022), 1026--1040. doi: 10.3934/mbe.2022047.

\bibitem{vobj_1}
R. Agarwal, T. Reed, How to end the COVID-19 pandemic by March 2022, SSRN, (2021). doi: 10.2139/ssrn.3826499.

\bibitem{vobj_2}
A. B. Gumel, E. A. Iboi, C. N. Ngonghala, G. A. Ngwa, Towards achieving a vaccine-derived herd immunity threshold for COVID-19 in the U.S., medRxiv, (2021). doi: 10.1101/2020.12.11.20247916.

\bibitem{vobj_3}
K. Kadkhoda, Herd Immunity to COVID-19: Alluring and Elusive, Amer. J. Clin. Pathol., 155 (4) (2021), 471--472. doi: 10.1093/AJCP/AQAA272.

\bibitem{vobj_4}
A. Fontanet, S. Cauchemez, COVID-19 herd immunity: Where are we?, Nat. Rev. Immunol., 20 (2020), 583--584. doi: 10.1038/s41577-020-00451-5.

\bibitem{vobj_5}
H. Liu, J. Zhang, J. Cai, X. Deng, C. Peng, X. Chen, et al., Herd immunity induced by COVID-19 vaccination programs to suppress epidemics caused by SARS-CoV-2 wild type and variants in China, medRxiv, (2021). doi: 10.1101/2021.07.23.21261013.

\bibitem{v_mask}
M. Shen, J. Zu, C. K. Fairley, J. A. Pagan, L. An, Z. Du, et al., Projected COVID-19 epidemic in the United States in the context of the effectiveness of a potential vaccine and implications for social distancing and face mask use, Vaccine, 39 (16) (2021), 2295--2302. doi: 10.1016/j.vaccine.2021.02.056.

\bibitem{ref_omae_aims2}
Y. Omae, Y. Kakimoto, J. Toyotani, K. Hara, Y. Gon, H. Takahashi, SIR model-based verification of effect of COVID-19 contact-confirmation application (COCOA) on reducing infectors in Japan, Math. Biosci. Eng., 18 (5) (2021) 6506--6526. doi: 10.3934/mbe.2021323.

\bibitem{delta_rate}
Advisory Board for Countermeasures in Japan against COVID-19 (54th), Document 3-3 (p.120), \url{https://www.mhlw.go.jp/content/10900000/000840251.pdf}, accessed December 24, 2021.

\bibitem{vrate}
Our World in Data, Statistics and Research: Coronavirus (COVID-19) Vaccinations, \url{https://ourworldindata.org/covid-vaccinations}, accessed December 24, 2021.

\bibitem{delta}
Advisory Board for Countermeasures in Japan against COVID-19 (40th), Document 3-3 (p.84), \url{https://www.mhlw.go.jp/content/10900000/000796736.pdf}, accessed December 12,2021.

\bibitem{toyo}
Toyo Keizai Online covid-19 Task Team, Coronavirus Disease (COVID-19) Situation Report in Japan, \url{https://toyokeizai.net/sp/visual/tko/covid19/}, accessed December 12, 2021.

\bibitem{kobayashi}
G. Kobayashi, S. Sugasawa, H. Tamae, T. Ozu, Predicting intervention effect for COVID-19 in Japan: state-space modeling approach. Biosci. Trends, 14 (3) (2020), 174--181.

\bibitem{v_accept}
M. Snehota, J. Vlckova, K. Cizkova, J. Vachutka, H. Kolarova, E. Klaskova, et al., Acceptance of a vaccine against COVID-19-a systematic review of surveys conducted worldwide. Bratislavske Lekarske Listy, 122 (8) (2021), 538--547. doi: 10.4149/bll\_2021\_086.

\bibitem{vcases}
Our World in Data, Statistics and Research: Coronavirus (COVID-19) Cases, \url{https://ourworldindata.org/covid-cases}, accessed January 5, 2022.

\bibitem{omic}
N. Andrews, J. Stowe, F. Kirsebom, S. Toffa, T. Rickeard, E. Gallagher, et. al., Effectiveness of COVID-19 vaccines against the Omicron (B.1.1.529) variant of concern, medRxiv (2021). doi: 10.1101/2021.12.14.21267615.
\end{thebibliography}
\end{document}